\documentclass{article}
\usepackage{emulateapj,psfig}
\usepackage{graphicx}
\usepackage{timesfonts}
\def\ltsima{$\; \buildrel < \over \sim \;$}
\def\lsim{\lower.5ex\hbox{\ltsima}}
\def\loe{\lower.5ex\hbox{\ltsima}}
\def\gtsima{$\; \buildrel > \over \sim \;$}
\def\gsim{\lower.5ex\hbox{\gtsima}}
\def\goe{\lower.5ex\hbox{\gtsima}}
\def\ergs{\rm \ erg \, s^{-1}}

\def\mdot {\dot M}

\def\msole {~M_{\odot}}

\lefthead{Campana et al.}
\righthead{Aql X-1 from outburst to quiescence}

\begin{document}

\title{Aquila X-1 from outburst to quiescence: 
the onset of the propeller effect and signs of a turned-on 
rotation-powered pulsar}

\authoremail{campana@merate.mi.astro.it}

\author{S. Campana\altaffilmark{1,2}, L. Stella\altaffilmark{3,2}, 
S. Mereghetti\altaffilmark{4}, M. Colpi\altaffilmark{5}, 
M. Tavani\altaffilmark{6,4,2}, D. Ricci\altaffilmark{7,2}, 
D. Dal Fiume\altaffilmark{8}, T. Belloni\altaffilmark{9}}

\affil{1. Osservatorio Astronomico di Brera, Via Bianchi 46, I-23807
Merate (Lc), Italy}
\affil{2. Affiliated to the International Center for Relativistic
Astrophysics (I.C.R.A.)}
\affil{3. Osservatorio Astronomico di Roma, Via dell'Osservatorio 2,
I-00040 Monteporzio Catone (Roma), Italy}
\affil{4. Istituto di Fisica Cosmica ``G.P.S. Occhialini'' del C.N.R., 
Via Bassini 15, 
I-20133 Milano, Italy}
\affil{5. Dipartimento di Fisica, Universit\`a degli Studi di Milano,
Via Celoria 16, I-20133 Milano, Italy}
\affil{6. Columbia Astrophysics Laboratory, Columbia University,
NY 10027 New York, USA}
\affil{7. BeppoSAX Science Data Center, Via Corcolle 19, I-00131 
Roma, Italy}
\affil{8. Istituto di Tecnologie e Studio delle Radiazioni 
Extraterrestri del C.N.R., Via Gobetti 101, I-40129 Bologna, Italy}
\affil{9. Astronomical Institute ``Anton Pannekoek'', University of 
Amsterdam and Center for High-Energy Astrophysics,
Kruislaan 403, NL-1098 SJ Amsterdam, The Netherlands}

\begin{abstract}
We report on the March-April 1997 BeppoSAX observations of Aql X-1, 
the first to monitor the evolution of the spectral and time variability 
properties of a neutron star soft X--ray transient from the outburst 
decay to quiescence.
We observed a fast X--ray flux decay, which brought the source luminosity 
from $\sim 10^{36}$ to $\sim 10^{33}\ergs$ in less than 
10 days. The X--ray spectrum showed a power law high energy tail with photon 
index $\Gamma\sim 2$ which hardened to $\Gamma \sim 1-1.5$ as the source 
reached quiescence. 
These observations, together with the detection by RossiXTE of a periodicity 
of a few milliseconds during an X--ray burst, likely indicate that the rapid 
flux decay is caused by the onset of the propeller effect arising from the very 
fast rotation of the neutron star magnetosphere. The X--ray luminosity and 
hard spectrum that characterise the quiescent emission can be consistently 
interpreted as shock emission by a turned-on rotation-powered pulsar.
\end{abstract}


\section{Introduction} 
Soft X--Ray Transients (SXRTs), when in outburst, show 
properties similar to those of persistent Low Mass X--Ray Binaries
containing a neutron star (LMXRBs; White et al. 1984; Tanaka \& 
Shibaza\-ki 1996; Campana et al. 1998).
The large variations in the accretion rate that are characteristic of SXRTs
allow the investigation of a variety of regimes for the neutron 
stars in these systems which are inaccessible to persistent LMXRBs. 
While it is clear that, when in outbursts, SXRTs are powered by accretion, 
the origin of the low luminosity X--ray emission that has been detected
in the quiescent state of several SXRTs is still unclear.
An interesting possibility is that a millisecond radio pulsar (MSP) turns on 
in the quiescent state of SXRTs (Stella et al. 1994). This would provide a
``missing link" between persistent LMXRBs and recycled MSPs. 
\begin{figure*}[!htb]
\psfig{figure=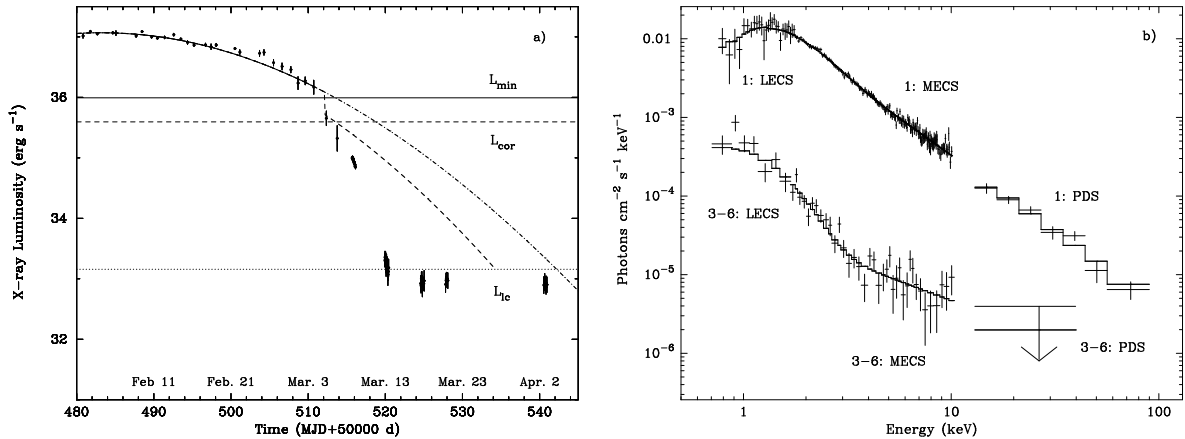,width=4.3cm}
\caption{
Light curve of the Feb.-Mar. 1997 outburst of Aql X-1 (panel {\it a}. 
Data before and after MJD 50514 were collected with the RossiXTE ASM 
(2--10 keV) and the BeppoSAX MECS (1.5--10 keV), respectively.
RossiXTE ASM count rates are converted to (unabsorbed) luminosities using 
a conversion factor of $4\times 10^{35}\ergs$ (before MJD 50512) and 
$2\times10^{35}\ergs$ (after MJD 50512) as derived from RossiXTE spectral fits
(Zhang, Yu \& Zhang 1998).
BeppoSAX luminosities are derived directly from the spectral data (see text).
The evolution of the flux from MJD 50480 to MJD 50512 is well fit 
by a Gaussian centered on MJD 50483.2. This fit however does not provide 
an acceptable description for later times (see the dot-dashed line), not even
if the accretion luminosity is calculated in the propeller regime 
(dashed line). 
The straight solid line represents the X--ray luminosity corresponding to
the closure of the centrifugal barrier $L_{\rm min}$
(for a magnetic field of $10^8$ G and a spin period of 1.8 ms)
and the straight dashed line the luminosity gap due to the action of the
centrifugal barrier, $L_{\rm cor}$. The dotted line marks the minimum 
luminosity in the propeller regime ($L_{\rm lc}$).
Panel {\it b} shows the BeppoSAX unfolded spectra of Aql X-1 during the early 
stages of the fast decline (1) and during the quiescent phase (3--6, summed). 
The best fit spectral model (black body plus power law) is
superposed to the data.}
\label{tot}
\end{figure*}
Aql X-1 is the most active SXRT known: more than 30
X--ray and/or optical outbursts have been detected so far. 
The companion star has been identified with the K1V variable 
star V1333 Aql and an orbital period of 19 hr has been measured
(Chevalier and Ilovaisky 1991).
The outbursts of Aql X-1 are generally characterised by a fast rise
(5--10 d) followed by a slow decay, with an $e-$folding time 
of 30--70~d (see Tanaka \& Shibazaki 1996 and Campana et al. 1998 and
references therein).
Type I X--ray bursts were discovered during the declining phase of an outburst 
(Koyama et al. 1981), testifying to the presence of a neutron star. 
Peak X--ray luminosities are in the $\sim (1-4)\times 10^{37}\ergs$ 
range (for the $\sim 2.5$ kpc distance inferred from its optical 
counterpart; Thorstensen et al. 1978). 
Close to the outburst maximum the X--ray spectrum is soft with an equivalent  
bremsstrahlung temperature of $k\,T_{\rm br} \sim 4-5$~keV. 
Sporadic observations of Aql X-1 during the early part of the  
outburst decay (Czerny et al. 1987; Tanaka \& Shibazaki 
1996; Verbunt et al. 1994) showed that when the source luminosity drops 
below $\sim 10^{36}\ergs$ the spectrum changes  
to a power law with a photon index of $\Gamma\sim 2$, 
extending up to energies of $\sim 100$ keV (Harmon et al. 1996).
ROSAT PSPC observations revealed Aql X-1 in quiescence on three occasions 
at a level of $\sim 10^{33}\ergs$ (0.4--2.4 keV; Verbunt et al. 1994). 
In this lower energy band the spectrum is considerably softer  and consistent 
with a black body temperature of $k\,T_{\rm bb} \sim 0.3$~keV.
\section{X--ray observations}
An outburst from Aql X-1 reaching a peak luminosity of $\sim 10^{37}\ergs$ 
(2--10 keV) was discovered and monitored starting from mid-February, 
1997 with the RossiXTE All Sky Monitor (ASM; Levine et al. 1997). 
Six observations were carried out with the BeppoSAX 
Narrow Field Instruments (NFIs) starting from March 8$^{th}$, 1997 
(see Table 1), with the aim of studying the final stages 
of the outburst decay. Fig. 1a shows the light curve of the Aql X-1 outburst
as observed by the RossiXTE ASM and BeppoSAX MECS. The first part of the 
outburst can be fit by a Gaussian with sigma $\sim 17$~d. This is 
not uncommon in SXRTs (e.g. in the case of 4U 1608--52; 
Lochner \& Roussel-Dupr\`e 1994).

The flux decay rate changed dramatically around MJD 50512 (March 5$^{th}$, 
1997). 
At the time of the first BeppoSAX observation (which started on 
March 8$^{th}$, 1997) the source luminosity was decreasing very rapid\-ly, 
fading by about 30\% in 11 hr, from a maximum level of $\sim 
10^{35}\ergs$. 
The second observation took place on March 12$^{th}$, 1997 when 
the source, a factor of $\sim 50$ fainter on average, reduced its flux 
by about 25\% in 12 hr. 
In the subsequent four observations the source luminosity attained a 
constant level of $\sim 6\times10^{32}\ergs$, consistent with previous 
measurements of the quiescent luminosity of Aql X-1 (Verbunt et al. 1994).
The sharp decrease after MJD 50512 is well described by an exponential 
decay with an $e-$folding time $\sim 1.2$ d.

The BeppoSAX LECS, MECS and PDS spectra during the fast decay phase, as well 
as those obtained 
by summing up all the observations pertaining to quiescence, can be fit 
with a model consisting of a black body plus a power law (see Table 2 and
Fig. 1b). The soft black body component remained nearly constant in temperature 
($kT_{\rm bb} \sim 0.3-0.4$ keV), but its radius decreased by a factor of 
$\sim 3$ from the decay phase to quiescence. The equivalent radius 
in quiescence ($R_{\rm bb}\sim 1$~km) is consistent with the ROSAT 
results (Verbunt et al. 1994).
The power law component changed substantially from 
the decay phase to quiescence: during the decay the photon index was 
$\Gamma \sim 2$, while in quiescence it hardened to $\Gamma\sim 1$. 
The two values are different with $> 90\%$ confidence (see Table 1).
The ratio of the 0.5--10 keV luminosities in the 
power law and black body components decreased by a factor of five between 
the first BeppoSAX observation and quiescence.
\section{Discussion}
\begin{table*}
\begin{center}
\caption{Summary of SAX NFIs observations.}
\label{tab1}
\begin{tabular}{cccccc}
Obs./Date               & LECS/MECS-PDS  & LECS Count Rate              & MECS Count Rate & PDS Count Rate\\
                        & Expos. (s)     & (c s$^{-1}$)                 & (c s$^{-1}$)    & (c s$^{-1}$)\\
\hline
1/March  8$^{th}$, 1997 &\, 5240/21342   & $0.84 \pm 0.02$              & $2.2\pm 0.01$                 & $0.87\pm0.06$ \\
2/March 12$^{th}$, 1997 &\, 3247/21225   & $(2.5\pm0.4) \times 10^{-2}$ & $(7.4\pm 0.2) \times 10^{-2}$ & $\lsim 0.19$ \\
3/March 17$^{th}$, 1997 &\, 5755/17258   & $(5.6\pm1.7) \times 10^{-3}$ & $(1.3\pm 0.1) \times 10^{-2}$ & $\lsim 0.24$ \\
4/March 20$^{th}$, 1997 &\, 4287/22589   & $(5.8\pm1.9) \times 10^{-3}$ & $(1.6\pm 0.1) \times 10^{-2}$ & $\lsim 0.17$\\
5/April 2$^{nd}$, 1997 &\, 8440/23576     & $(6.2\pm1.3) \times 10^{-3}$ & $(1.2\pm 0.1) \times 10^{-2}$& $\lsim 0.17$ \\
6$^{\rm o}$/May 6$^{th}$, 1997 & 11789/21703 & $(6.7\pm1.1) \times 10^{-3}$ &   --- & $\lsim 0.19$ \\
\hline
\end{tabular}
\end{center}
{\noindent $^{\rm o}$ No MECS data were obtained.} 
\end{table*}
\begin{table*}
\begin{center}
\caption{Summary of spectral fits.}
\label{tab2}
\begin{tabular}{ccccccc}
Obs.$^*$ & Black body  & Black body  & Power law    & PL/BB     & Mean Luminosity$^\dag$  & Reduced\\ 
         & $k\,T_{\rm bb}$ (keV) & $R_{\rm bb}$ (km)& $\Gamma$ & flux ratio$^{\ddag}$&(erg s$^{-1}$)& $\chi^2$\\
\hline
1    & $0.42\pm0.02$&$2.6\pm0.3$&$1.9\pm0.1$& 3.7 & $9\times 10^{34}$ &1.0 \\ 
2    & $0.3_{-0.2}^{+0.1}$&$0.7_{-0.3}^{+6.6}$&$1.8\pm0.7$& 1.6 & $2\times 10^{33}$ &0.9 \\ 
\ 3--6 & $0.3\pm0.1$&$0.8_{-0.1}^{+0.4}$&$1.0\pm0.3$& 0.7 & $6\times 10^{32}$ &1.3 \\
\hline
\end{tabular}
\end{center}
{
\noindent Errors are $1\,\sigma$. 

\noindent $^*$ Spectra from the LECS and MECS (and PDS for the first 
observation) detectors have been considered.
The spectra corresponding to the quiescent state have 
been summed up, in order to increase the statistics and an upper limit 
from the summed PDS data was also used to better constrain the 
power law slope.

\noindent $\dag$ Unabsorbed X--ray luminosity in the energy range 
0.5--10 keV. In the case of the first observation the PDS data 
were included in the fit (the unabsorbed 0.5--100 keV luminosity 
amounts to $2\times 10^{35}\ergs$).

\noindent $\ddag$ Power law to black body flux ratio in the 0.5--10 keV 
energy range.}

\end{table*}
The BeppoSAX observations enabled us to follow for the first time the 
evolution of a SXRT outburst down to quiescence. 
The sharp flux decay leading to the quiescent state of Aql X-1 
is reminiscent of the final evolution of dwarf novae outbursts
(e.g.  Ponman et al. 1995; Osaki 1996), although there are obvious differences 
with respect to the X--ray luminosities and spectra involved in the 
two cases, likely resulting from the different efficiencies 
in the gravitational energy release between white dwarfs and neutron stars.
 
Models of low mass X--ray transient outbursts hosting an old 
neutron star or a black hole are largely built in analogy with dwarf
novae outbursts. In particular, 
van Paradijs (1996) showed that the different range of time-averaged 
mass accretion rates over which the dwarf nova and low mass X--ray transient 
outbursts were observed to take place is well explained by the higher 
level of disk irradiation caused by the higher accretion efficiency 
of neutron stars and black holes. 
However, the outburst evolution of low mass X--ray transients 
presents important differences. In particular, the steepening in the X--ray  
flux decrease of Aql X-1 has no clear parallel in low mass X--ray transients 
containing Black Hole Candidates (BHCs). The best sampled light curves 
of these sources show an exponential-like decay (sometimes with a superposed 
secondary outburst) with an $e-$folding time of $\sim 30$ d and
extending up to four decades in flux, with no indication of a sudden 
steepening (Chen et al. 1997). In addition, BHC transients display 
a larger luminosity range between outburst peak and quiescence than 
neutron star SXRTs (Garcia et al. 1998 and references therein). 
Being the mass donor stars and the binary parameters quite similar in the 
two cases, it appears natural to attribute these differences to the  
different nature of the underlying object:
neutron stars possess a surface and, likely, a magnetosphere, while BHCs do 
not. 

When in outburst accretion down to the neutron star surface takes place 
in SXRTs, as testified by the similarity of
their properties with those of persistent LMXRBs, especially
the occurrence of type I bursts and the X--ray spectra and luminosities. 
The mass inflow rate during the outburst decay decreases, causing 
the expansion of the magnetospheric radius, $r_{\rm m}$.
Thus, accretion onto the neutron star surface can continue as long as the 
centrifugal drag exerted by the corotating magnetosphere on the accreting 
material is weaker than gravity (Illarionov \& Sunyaev 1975; Stella et al. 
1986).
This occurs above a limiting luminosity  $L_{\rm min}=G\,M\,\mdot_{\rm min}/R 
\sim 4\times10^{36} \,B_8^2\,P_{-3}^{-7/3} ~{\rm erg\,s^{-1}}$,
where $G$ is the gravitational constant; $M$, $R$, $B=B_8\,10^8$ G and 
$P=P_{-3}\,10^{-3}$ ms are the neutron star  mass, radius, magnetic 
field and spin period, respectively (here and in the following we assume  
$M=1.4\msole$ and $R=10^6$ cm). As $r_{\rm m}$ reaches the corotation
radius, $r_{\rm cor}$, accretion onto the surface is inhibited
and a lower accretion luminosity ($<L_{\rm {min}}$) of 
$L_{\rm cor}=G\,M\,\mdot_{\rm min}/r_{\rm cor}\sim 2\times 
10^{36}\,B_8^2\,P_{-3}^{-3}\ergs$ is released. After this luminosity 
gap the source enters the propeller regime.
If the mass inflow rate decreases further,
the expansion of $r_{\rm m}$ can continue up to the light cylinder radius, 
$r_{\rm lc}$, providing a lower limit to the accretion luminosity that 
can be emitted in the propeller regime
$ L_{\rm lc}=G\,M\,\mdot_{\rm lc}/r_{\rm lc}\sim 2\times 
10^{34}\,B_8^2\,P_{-3}^{-9/2}\ergs$.
Below $L_{\rm {lc}}$ the radio pulsar mechanism may turn on and the pulsar 
relativistic wind interacts with the incoming matter pushing it 
outwards. Matter inflowing through the Roche lobe is stopped by the 
radio pulsar radiation pressure, giving rise to a shock front (Illarionov \& 
Sunyaev 1975; Shaham \& Tavani 1991).
Clearly these regimes have no equivalent in the case of black hole accretion. 
\subsection{The onset of the propeller}
During the February-March 1997 outburst of Aql X-1, RossiXTE observations 
led to the discovery of a nearly coherent modulation at 
$\sim 550$~Hz ($\sim 1.8$ ms) during a type I X--ray burst. A single  
QPO peak, with a centroid frequency ranging from $\nu_{QPO}\sim 750$ to 
830 Hz, was also observed at two different flux levels, when the persistent 
luminosity was $\sim 1.2\times 10^{36}\ergs$ and $\sim 1.7\times 10^{36}\ergs$ 
(Zhang et al. 1998). 
In the presence of a single QPO peak, the magnetospheric and sonic point 
beat frequency models (Alpar \& Shaham 1985;
Miller et al. 1997) interpretation 
is ambiguous in that the QPO peak could represent either the 
Keplerian frequency at the inner disk boundary 
or the beat frequency. Moreover, the burst periodicity at 
$\sim 550$~Hz may represent the neutron star spin frequency, $\nu_s$, or 
half its value (Zhang et al. 1998). In either cases, the possibility 
that accretion onto the 
neutron star surface takes place even in the quiescent state of 
Aql X-1 faces serious difficulties: 
for a quiescent luminosity of order $10^{33}\ergs$ a magnetic field
of only $\lsim 5\times 10^6$ G would be required, in order
to fulfill the condition $r_{\rm m}\lsim r_{\rm cor}$.
For such a low magnetic field, Aql~X-1 and, by inference, LMXRBs 
with kHz QPOs can hardly be the progenitors of recycled MSPs. 
More crucially, the marked steepening in the outburst decay that takes 
place below  $\sim 1\times 10^{36}\ergs$, is accompanied by a marked 
spectral hardening, resulting from a sudden decrease of the 
flux in the black body spectral component. This is clearly suggestive of a 
transition taking place deep in the gravitational well of the neutron star, 
where most of the X--rays are produced.
The most appealing mechanism is a transition to the propeller regime, 
where most of the inflowing matter 
is stopped at the magnetospheric boundary (Zhang, Yu \& Zhang 1998).
In Fig. 1a, the luminosity at MJD 50512 is identified with $L_{\rm {min}}$ 
and the lower horizontal lines indicate the luminosity interval during 
which Aql X-1 is likely in the propeller regime.

Additional information on the neutron star magnetic field (and spin) can be 
inferred as follows. The observed ratio of the luminosity, 
$L_{QPO}$, when the QPO at 
$\sim 800$~Hz were detected and the luminosity $L_{\rm min}$ 
when the centrifugal barrier closes is $L_{QPO}/L_{\rm min} \sim 1.2-1.7$. 
At $L_{\rm min}$ the Keplerian frequency of 
matter at the magnetospheric boundary is, by definition, equal to 
the spin frequency, i.e.  $\nu_s \sim 550$ or $\sim 275$~Hz
for Aql X-1. Based on beat-frequency models, at $L_{QPO}$ the Keplerian 
frequency at the 
inner disk boundary can be either $\nu_{K,QPO}\sim 800$~Hz or 
$\nu_{K,QPO}\sim \nu_s + 800$~Hz, depending on whether the single 
kHz QPOs observed corresponds to the Keplerian or the beat frequency.
In the magnetospheric beat-frequency models, simple theory predicts that 
the Keplerian 
frequency at the magnetospheric boundary is $\propto L^{3/7}$; 
in the radiation pressure-dominated regime relevant to the case at hand, 
the Ghosh and Lamb (1992) model predicts instead $\propto L^{3/13}$. 
Therefore we expect $\nu_{K,QPO}/\nu_s \sim (L_{QPO}/L_{\rm min})^{3/7} 
\sim 1.2$ and $\nu_{K,QPO}/\nu_s \sim (L_{QPO}/L_{\rm min})^{3/13} 
\sim 1.1$, in the two models, respectively. Such a low ratio clearly favors 
the interpretation in which $\nu_{K,QPO}\sim 800$~Hz and $\nu_s\sim 550$~Hz. 
In the sonic point beat-frequency model (Miller et 
al. 1997), the innermost disk radius is well within the 
magnetosphere, implying that the Keplerian frequency at the magnetospheric 
boundary is $< \nu_{K,QPO}$. In this case all possible combinations 
of $\nu_s$ and $\nu_{K,QPO}$ are allowed. 
By using the observed $L_{\rm min}$, a neutron star magnetic field of 
$B\sim 1-3 \times 10^8$~G (depending on the adopted model of the 
disk-magnetosphere interaction) is obtained in the case $\nu_s\sim 550$~Hz 
and $B\sim 2-6 \times 10^8$~G in the case $\nu_s\sim 275$~Hz. 

Once in the propeller regime, the accretion efficiency decreases further 
as the magnetosphere expands for decreasing mass inflow rates
($L_{\rm acc}\propto \mdot^{9/7}$). The $\sim 1$~d exponential-like luminosity 
decline observed with BeppoSAX is considerably faster than the propeller 
accretion luminosity extrapolated from the first part of 
the outburst (e.g. the Gaussian profile shown by the dashed line in Fig. 1a). 
We note here that the spectral transition accompanying the onset of the
centrifugal barrier may also modify the irradiation properties
of the accretion disk, contributing to X--ray luminosity turn off.
Alternatively, an active contribution of the ``propeller'' mechanism or 
the neutron star spin-down energy dissipated into the inflowing
matter cannot be excluded. 
\subsection{A turned-on rotation-powered pulsar?}
It is unlikely that the quiescent luminosity of Aql~X-1 is powered by 
magnetospheric accretion in the propeller regime. As shown in Fig. 1a, 
the quiescent X--ray luminosity is probably lower than the minimum 
magnetospheric accretion luminosity $L_{\rm lc}$ allowed in the propeller 
phase (this remains true for $B\gsim 6\times 10^7$ G if $\nu_s\sim 550$ Hz,
and for $B\gsim 3\times 10^8$ G if $\nu_s\sim 275$ Hz).
Moreover the BeppoSAX X--ray spectrum 
shows a pronounced decrease in the power law to black body flux ratio 
together with a flattening of the power law component between the fast 
decay phase and quiescence, suggesting that a  
transition to shock emission from the 
interaction of a radio pulsar wind with the matter outflowing from 
the companion star has taken place.  
Note that an  X--ray spectrum with a slope of $\Gamma \sim 1.5$ has been 
observed from the radio pulsar PSR~B1259--63 immersed in the wind of its
Be star companion. Models of this interaction predict that a 
power law X--ray spectrum with a slope around $\Gamma \sim 1.5$ should be 
produced for a wide range of parameters (Tavani \& Arons 1997).
 
The additional soft X--ray component observed during the outburst 
decay (see Table 2) might be emitted at the polar caps as a result of the 
residual neutron star accretion in the propeller phase.
Note that the equivalent black body radius decreases for
decreasing X--ray luminosities, just as it would be expected if the 
magnetospheric boundary expanded.
Alternatively, the black body-like spectral component observed 
in quiescence could be due to the stre\-aming of energetic particles that hit  
the polar caps, in close analogy to the soft X--ray 
component observed, in MSPs, at the weaker level of $\sim 10^{30}-10^{31}\ergs$ 
(Becker \& Tr\"umper 1997).

Assuming a magnetic field in the range derived in section 3.1 
(i.e. $B\sim 1-3\times10^8$ G for $\nu_s=550$ Hz and $B\sim 2-6\times10^8$ 
G for $\nu_s=275$ Hz), we can consistently explain the $\sim 10^{33}\ergs$ 
quiescent X--ray luminosity as powered by a radio pulsar enshrouded by 
matter outflowing from the companion star, if the conversion efficiency 
of spin-down luminosity to X--ray is $\sim 0.1-10$\%. 
This is consistent with modeling and observations of enshrouded pulsars
(Tavani 1991; Verbunt et al. 1996).

There are chances of observing a MSP (a simple scaling from MSPs implies 
a signal at 400 MHz of $\sim 10$ mJy; see Kulkarni et al. 1990), 
even though the emission would probably be sporadic, like in the case 
of the pulsar PSR~1744--24A due to the large amount of circumstellar 
matter (see Lyne et al. 1991; Shaham \& Tavani 1991).

In summary Aql~X-1 appears to provide the first example of an 
old fast rotating neutron stars undergoing a transition to the propeller
regime at first, followed by a transition to the radio pulsar 
regime, as the transient X--ray emission approaches its quiescent level.    
Therefore, Aql X-1 (and possibly SXRTs in general) likely represents
the long-sought ``missing link'' between LMXRBs and recycled MSPs. 

\acknowledgments
We thank J. van Paradijs, F.K. Lamb and an anonymous referee for useful 
comments. S.N. Zhang, W. Yu and W. Zhang shared with us their RossiXTE 
results in advance of publication.

\end{document}